\documentclass[aps,pra,twocolumn]{revtex4}

\usepackage{graphicx}
\usepackage{color}
\usepackage{bm}
\usepackage{upgreek}
\usepackage{amsmath, amsthm, amssymb,mathrsfs}
\usepackage{csquotes}
\usepackage[colorlinks=true, citecolor=blue,allcolors=blue]{hyperref}

\begin{document}

\title{Dynamics of Tunneling Ionization using Bohmian Mechanics}

\author{Nicolas Douguet$^{1,2}$ and Klaus Bartschat$^1$}

\affiliation{$^1$Department of Physics and Astronomy, Drake University, Des Moines, Iowa 50311, USA}
\affiliation{$^2$Department of Physics, University of Central Florida, Orlando, 32816, Orlando, USA}

\date{\today}


\begin{abstract}
Recent atto\-clock experiments and theoretical studies regarding the strong-field ionization of 
atoms by \hbox{few-cycle} infrared pulses revealed new features that have attracted much attention.
Here we investigate tunneling ionization and the dynamics of the electron probability
using Bohmian Mechanics. We consider a one-dimensional problem 
to illustrate the underlying mechanisms of the ionization process.
It is revealed that in the major part of the below-the-barrier ionization regime, in an intense and short infrared pulse,
the electron does not tunnel ``through'' the entire barrier, 
but rather already starts from the classically forbidden region. 
Moreover, we high\-light the correspondence
between the probability of locating the electron at a particular initial position and its asymptotic momentum. 
Bohmian Mechanics also provides a natural definition of mean tunneling 
time and exit position, taking account of the time dependence of the barrier. Finally,
we find that the electron can exit the barrier with significant kinetic energy, thereby corroborating the results of a recent study
[Camus {\it et al.}, Phys.\ Rev.\ Lett.~{\bf 119} (2017) 023201].
\end{abstract}

\maketitle

\section{Introduction}\label{sec:Intro}
The tunneling ionization of an electron in an ultra-short intense optical laser pulse represents a 
purely quantum process whose theoretical description remains challenging.
Numerous treatments, based on various approximations, have been elaborated to model ionization 
in the tunneling regime, e.g., using the adiabatic theorem \cite{Joachain00}, the strong-field 
approximation (SFA) \cite{Keldysh65,Faisal73,Reiss80}, the closed-orbit theory \cite{Du87}, the simple-man's model \cite{Corkum89}, 
as well as more recent techniques \cite{Klaiber15,Yakaboylu14,Torlina12,Torlina14}.
Although these models can already make impressive predictions, the ultra\-fast electron dynamics in a time-varying barrier
remains a process under great scrutiny that has been triggering extensive theoretical work. Improving our understanding of tunneling ionization is crucial 
for high-order harmonic generation (HHG), coherent quantum control, and atto\-second science in general. 

The present study is devoted to a description of tunneling ionization employing  
Bohmian Mechanics \cite{Bohm52,Durr09,Botheron10}, which has recently attracted much attention
\cite{Zimmermann16,Steinberg11,Braverman13,Steinberg16,Ivanov17,Song17,Benseny14,Jooya1,Jooya2,Jooya3,Jooya4}.
It will be shown that computing the streamlines of the wavefunction probability 
over time provides valuable insights and a natural route to understanding complex
ultra\-fast mechanisms. Despite the fact that Bohmian Mechanics leads to the same final results
as Quantum Mechanics, it offers an alternative  route to the complex time evolution of a wavepacket
by considering the streamlines of the electron probability over time while going beyond the SFA. 
Relating the wavefunction dynamics
to particle trajectories, as for instance done in the Feynman path integral approach or in semi\-classical models,
represents a very appealing aspect of Bohmian Mechanics.

An important topic to which Bohmian Mechanics can make a unique contribution concerns the understanding of
 ``tunneling time"  through a potential barrier, as recently considered in Ref.~\cite{Zimmermann16}. 
The concept of tunneling time (e.g., Larmor \cite{Buttiker83}, B\"{u}ttiker-Landauer \cite{Buttiker82}, Pollack-Miller \cite{Pollack84}, 
or Eisenbud-Wigner times \cite{Wigner55}) is a fuzzy concept, as
it cannot be obtained directly from
a physical observable. Since the various definitions lead to different
results \cite{Landsman14,Zimmermann16}, one might even question the relevance of
a tunneling time. On the other hand, the concept plays a central role 
in the Keldysh theory \cite{Keldysh65}, since it provides a criterion to separate the multi\-photon and the tunneling 
ionization regimes.

Revisiting the concept of tunneling time has become highly appropriate 
with the advent of atto\-clock experiments \cite{Landsman14,Camus17} and
debates around the claim of Torlina {\it et al.}~\cite{Torlina15} that optical tunneling in atomic hydrogen is instantaneous.
Indeed, two recent studies \cite{Ni16,Sainadh17}
obtained results supporting a tunneling ionization time close to zero,
while others~\cite{Landsman14,Zimmermann16,Camus17}
reported a nonzero tunneling time 
for traversing the barrier. Note that the tunneling ionization time defined in \cite{Torlina15,Ni16} corresponds 
to the moment at which the electron appears at the tunnel exit 
with respect to the instant of maximum field strength, and thus
it does not necessarily contradict the results of Refs.~\cite{Landsman14,Zimmermann16}. 
One might also suggest that part of the disagreement observed between different studies
is due to electron correlations in multi\-electron systems. However,
the recent work by Majety and Scrinzi \cite{Majety17} on helium revealed that electron correlations should have no effect 
on the asymptotic electron momentum offset angle. 

In this study, we show 
that Bohmian Mechanics provides natural definitions of tunneling ionization time, traversing time, and exit points for 
each trajectory, while explicitly accounting for the barrier dynamics. 
Bohmian Mechanics can provide a picture of the time propagation under a barrier 
without invoking imaginary tunneling time. Bohmian Mechanics might thus have the potential to reveal and rationalize 
the dynamics above and below a barrier, while only using familiar concepts.

In order to establish the basic ideas, we consider a one-dimensional model problem, which contains the principal ingredients 
of the tunneling process without adding non\-essential complexities. The study is thus not intended as a direct application
to a current experimental problem, but as the presentation of an alternative approach with attractive characteristics
for application in ultra\-fast physics, with the perspective that a similar approach could be performed on a realistic case.
Specifically, it will be shown that even for the one-dimensional problem considered in this work, 
Bohmian Mechanics provides important physical results that can hopefully be transferred to the real case.

Specifically, we demonstrate that the major part of below-the-barrier-ionization (BBI) induced
by an intense ultra\-short infrared pulse originates from the electron probability
initially located inside the classically forbidden region, i.e., from the tail of the initial ground-state wavefunction. 
Hence the picture of the probability flow traversing the entire barrier from the inner to the outer classically 
allowed regions is fundamentally flawed. Furthermore, we show that an above-threshold ionization (ATI) photo\-electron spectrum 
can be accurately reproduced through Bohmian Mechanics, thus leading to an appealing
correspondence between the total number of absorbed photons and the initial 
probability distribution of the electron position. Bohmian Mechanics also provides 
indications on when the quantum effects become negligible, 
i.e., when the so-called ``quantum force" \cite{Bohm52,Durr09} vanishes. 
As a result, a particle emerging from the barrier can still exhibit quantum behavior for a significant time.

This manuscript is organized as follows:
In the next section~\ref{sec:Theory}, we outline the theoretical approach and describe the model considered. 
Our results are presented and discussed in Sec.~\ref{sec:Results}, where we consider Bohmian trajectories, 
their correspondence to the final quantum photoelectron ionization spectrum, and how they can be used to define 
tunneling time and exit position. Section \ref{sec:conclusion} summarizes our conclusions.

Unless stated otherwise, atomic units are used throughout the manuscript.

\section{Theoretical approach}\label{sec:Theory}
Suppose $\varphi(x,t) = R(x,t)\exp{\left[iS(x,t)\right]}$,with $R$ and $S$ being real-valued functions, is the solution of the time-dependent
Schr\"{o}dinger Equation (TDSE). 
In Bohmian Mechanics in one dimension,
electron trajectories are computed from the following set of equations \cite{Bohm52,Ivanov17}:
\begin{eqnarray}
-\frac{\partial S(x,t)}{\partial t} &\!=\!& \frac{1}{2}\left(\frac{\partial S(x,t)}{\partial x}\right)^2 \!\!\!+ V_C(x,t) \!+\! V_Q(x,t); \label{eq:1}\\
\frac{\partial \rho(x,t)}{\partial t} &\!=\!& -\nabla\cdot[{\rho(x,t)v(x,t)}] \label{eq:2}.
\end{eqnarray}
Here $\rho(x,t) = R(x,t)^2$ is the probability density and
$v(x,t) = \Re\{[\hat p\,\varphi(r,t)]/\varphi(r,t)\}$ is the velocity field, where $\Re\{X\}$ denotes the real part of $X$ and $\hat p$
the momentum operator.  Furthermore, $V_C(x,t)$ and \hbox{$V_Q(x,t) = -0.5\Delta R(x,t)/R(x,t)$} are the classical 
and quantum potentials, respectively. Equation (\ref{eq:1}) is the Hamilton-Jacobi equation with the addition of
the quantum potential $V_Q(x,t)$ \cite{Riggs07}, while Eq.~(\ref{eq:2}) is the continuity equation for a current probability density 
$j(x,t) = \rho(x,t)v(x,t)$. Equations (\ref{eq:1}) and (\ref{eq:2}) are formally equivalent to the TDSE.

After evaluating the velocity field $v(x,t)$ from the solution $\varphi(x,t)$ of the TDSE, we integrate 
the equation $dx(t)/dt=v(x,t)$, starting from an initial position $x(x_0;0)=x_0$ at $t=0$, 
to compute classical trajectories $x(x_0;t)$ of the probability streamlines.
[Note that we maintain the commonly-used notation here, even though the time-dependent position $x(t)$ is not
the same as the spatial grid on which the various functions are defined.]  
These Bohmian trajectories have a broader significance than classical trajectories
and can, for instance, be used to reconstruct the wavefunction at any time \cite{Botheron10,Schleich13}. Very importantly,
the quantum potential allows Bohmian trajectories to penetrate into classically forbidden regions.


\begin{figure}[b]
\includegraphics[width=\columnwidth]{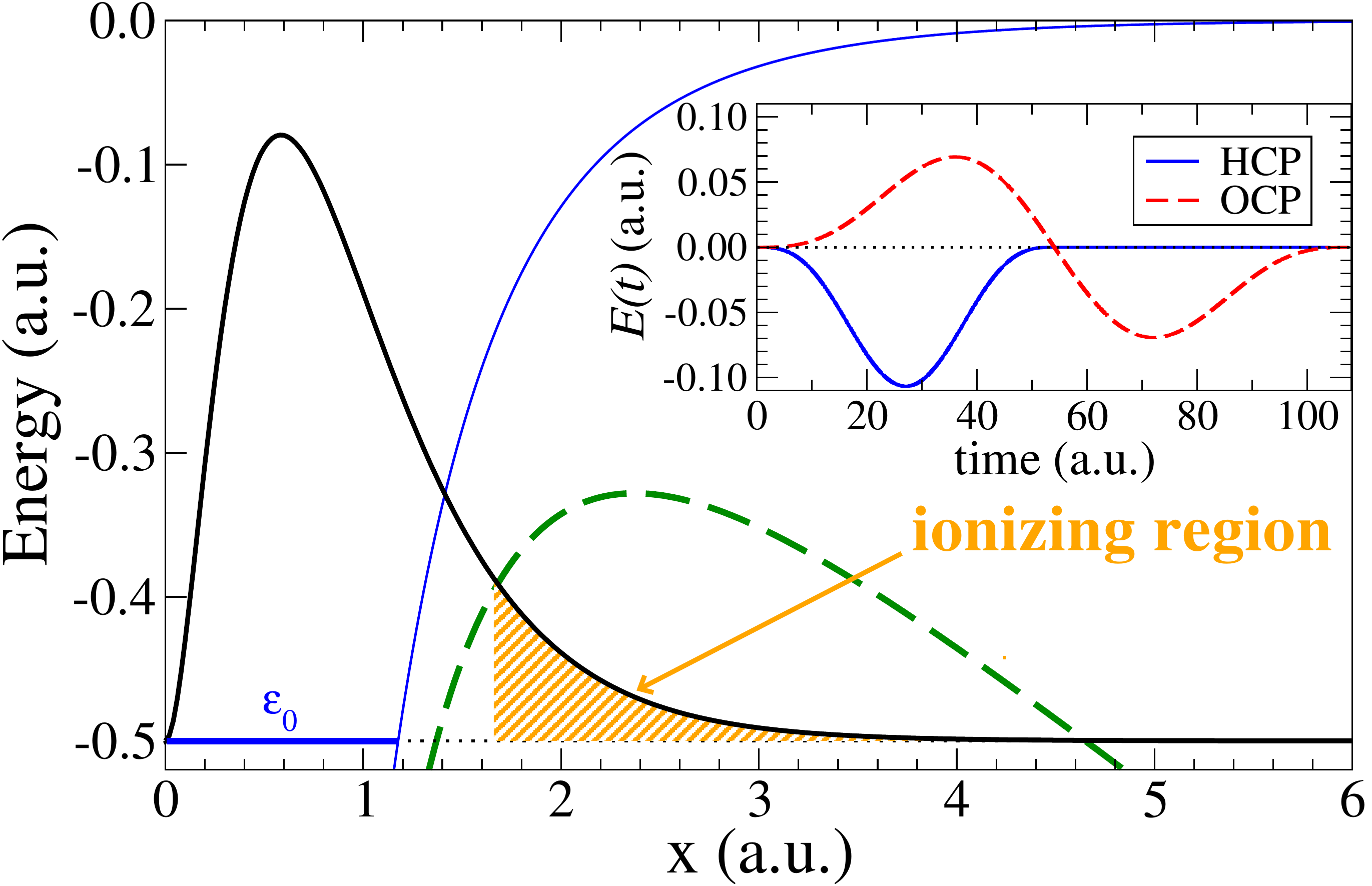}
 \caption{ 
 Representation of the ground-state probability (thick black line),
 the field-free Yukawa potential (thin blue line), and the classical potential at the maximum field strength (dashed green line)
 for $I_0 = 4\times 10^{14}~$W/cm$^2$.
 The insert shows the electric fields of our HCP and OCP (see text for details).}
 \label{fig:1}
\end{figure}

We consider a one-dimensional model, $x\in[-\infty,+\infty]$, and use, as our principal example, a short-range Yukawa potential \hbox{$V_0(x) = -Z\exp(-|x|)/|x|$}, 
 truncated at $\varepsilon=0.01$, such that $V_0(x) = V_0(\varepsilon)$ for $|x|<\varepsilon$.  This potential was also used in \cite{Abramovici09}.
 It could, for instance, model the photodetachment of an atomic anion. 
 We also carried out calculations for a truncated Coulomb potential, i.e., a one-dimensional H~atom, 
 and we discuss some of these results below.

Due to its illustrative advantages and regular behavior \cite{Abramovici09}, 
we consider the initial wavefunction $\varphi_0(x)$ to be the odd-parity eigenstate 
of lowest energy.  For $x\ge0$, therefore, $\varphi_0(x)$ is 
the (reduced) radial part of the $1s$ ground state of the corresponding three-dimensional problem. 
We set \hbox{$Z=-1.9083$} to produce an energy of \hbox{$\varepsilon_0=-0.5~$}.

The electric field $E(t) = E_0f(t)\sin(\omega t+\phi)$ has amplitude $E_0$, frequency $\omega = 0.05811$ ($\lambda=784$ nm), phase~$\phi$,
and a sine-squared envelope $f(t) = \sin^2(\Omega t)$, where $\Omega = \omega/2N$ with $N$ as the number of cycles. 
The TDSE is solved in the length gauge, with $V_{C}(x,t) = V_0(x) + E(t)x$, by a finite-difference method.

Throughout this study, we mostly consider a half-cycle pulse (HCP) with $\phi=180^\circ$, 
so that the electron is 
pulled towards the positive direction. 
This represents the simplest case to illustrate the essence of the BBI process. 
We choose a peak intensity
$I_0 = 4\times 10^{14}$ W/cm$^2$ 
to remain relatively far from 
over-the-barrier ionization (OBI) starting at $I_{\rm OBI} = 1.2\times10^{15}$ W/cm$^2$.
We also consider a one-cycle pulse (OCP) carrying the same energy as the HCP, i.e., with
$I_0 \approx 1.68\times10^{14}{\rm W/cm}^2$ and $\phi=0^\circ$. 
The Yukawa potential, $V_0(x)$, and the classical potential 
at the maximum field strength are plotted for $x\ge0$ in \hbox{Fig.~\ref{fig:1}}.

\section{Results and discussion}
\label{sec:Results}
\subsection{Bohmian dynamics} 

\begin{figure}[t]
\includegraphics[width=1.0\columnwidth]{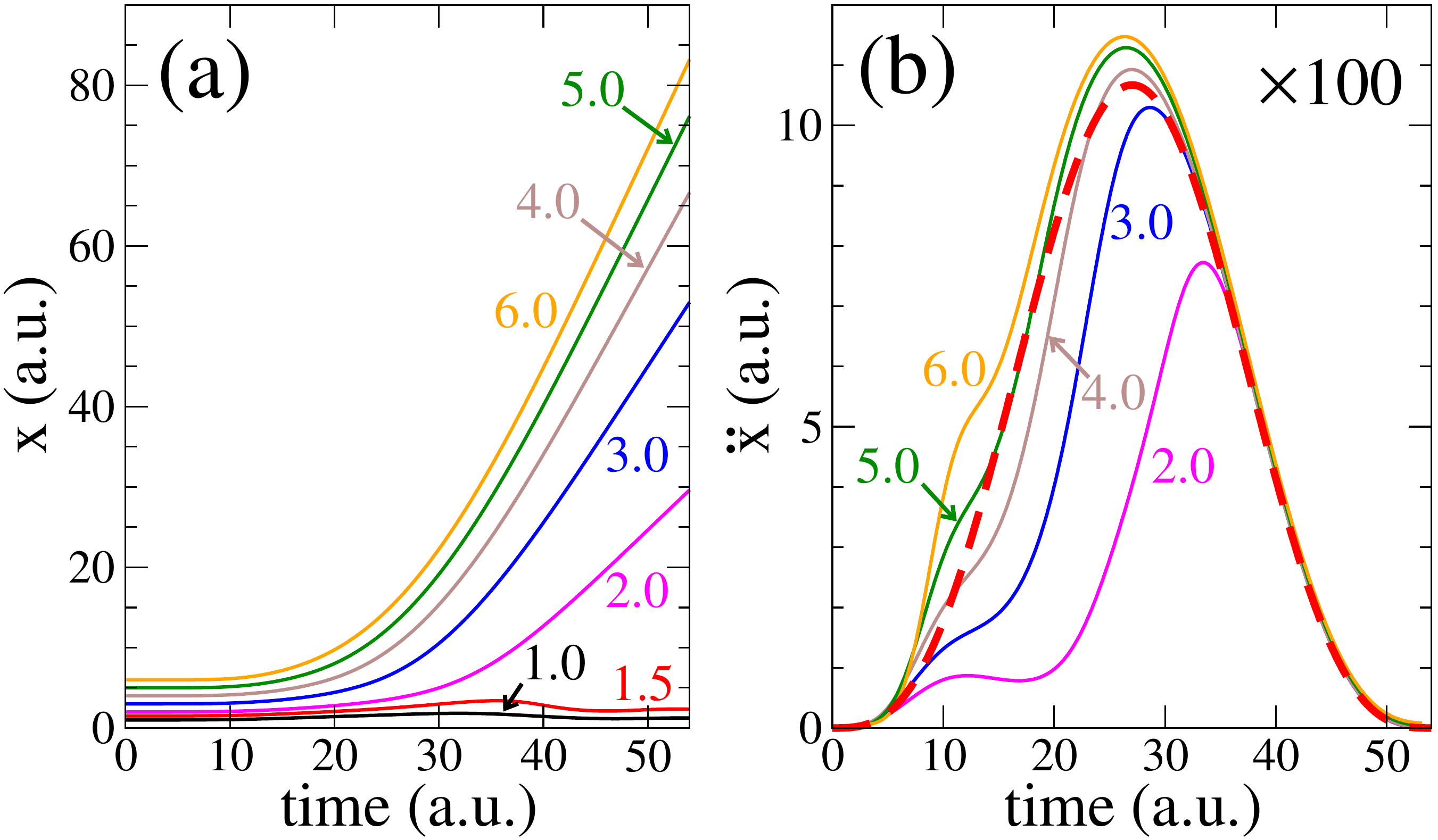}
 \caption{(Color online) Bohmian trajectories (a) and associated acceleration (b) as a 
 function of time labeled by their initial position $x_0$. 
The dashed red curve represents the time-dependent electric force $F_{\rm e}(t) = -E(t)$. }
 \label{fig:2}
\end{figure}

Having obtained the numerical solution of the TDSE, we computed thousands of trajectories using the Runge-Kutta method~\cite{NumRep}, 
starting from rest (\hbox{$j(x,0)=0$}) at initial positions~$x_0$. 
A few of these trajectories, plotted in \hbox{Fig.~\ref{fig:2}~(a)},
exhibit a smooth variation without apparent interference between paths. 
This simple situation contrasts with the cases of a few-cycle pulse \cite{Botheron10}, as well as the well-known example of 
entangled trajectories in the double-slit problem with photons~\cite{Steinberg11}. 

A rich amount of information can be extracted from the Bohmian trajectories.
First, we found that only trajectories starting in the ``ionizing region'' defined by $x_0\ge x_{\rm th}$ 
become asymptotically free, i.e., with a final speed \hbox{$v_\infty > 0$}. As seen in Fig.~\ref{fig:1}, 
this ionizing region is located inside the classically forbidden region.
It starts at $x_{\rm th} =1.65$~a.u., 
whereas the inner turning point 
for the energy $\varepsilon_0$ is $x_{\rm cl}=1.17~{\rm a.u.}$. At the lower intensity of $I_0 = 2\times 10^{14}~$W/cm$^2$,  
$x_{\rm th} =2.70$~a.u., deep inside the forbidden region, while $x_{\rm th} =x_{\rm cl}$ near $I_0 \approx 6\times 10^{14}~$W/cm$^2$.
Consequently, only for intensities close to the OBI regime, part of the tunneling can be considered as occurring across the barrier. 

Even though the characteristics of the ionizing region 
depend on the exact form of the pulse and potential, the latter trend survives for a Coulomb potential and a OCP.  
Given that OBI occurs at $I_{\rm OBI}=1.5\times 10^{14}~$W/cm$^2$
in atomic hydrogen, $x_{\rm th} =2.2$ at $I_0 = 10^{14}~$W/cm$^2$ while $x_{\rm cl}=2.0$. 
For the OCP, the ionizing region exists for $|x_0|$ above a certain threshold, which is different
for $x\ge0$ and $x<0$. 

We then repeated the calculations for 2- and 3-cycle pulses and 
confirmed these general findings. The trend is expected to also hold in three dimensions, which would confirm
the conclusion expressed in Ref.~\cite{Khausal17}, namely: In a specific ionization regime, tunneling only occurs from the tail
of the initial wavefunction.

In order to understand the tunneling dynamics, it is instructive to look at the acceleration $\ddot x(x_0;t)$ 
of the trajectories as a function of time. \hbox{Figure~\ref{fig:2}~(b)} shows the acceleration of a few
asymptotically free trajectories, as well as the electric force $F_{\rm e}(t)$ created by the HCP. Not surprisingly, 
the acceleration approaches $F_{\rm e}(t)$ with increasing time, and the acceleration for small initial positions $x_0$ 
merges the latter onto the electric force. The discrepancy between the acceleration and $F_{\rm e}(t)$ is
partly due to the Yukawa force $F_0(x) = -dV_0(x)/dx$, but also to the quantum force 
$F_Q(x,t) = -\partial V_Q(x,t)/\partial x$. Initially, $F_Q(x,0) = -F_0(x)$, since the system is
in equilibrium at zero field. For trajectories starting at large $x_0$, $F_0(x)$ 
becomes rapidly negligible in comparison with $F_Q(x,t)$ due to the short-range nature of the potential. 
The quantum force $F_Q(x,t)$ provides the extra energy that,
in conjunction with the field energy, allows the particle to cross the barrier and emerge into the classically allowed region. 
It is also interesting to note the counterintuitive fact that some trajectories, while still under the barrier, can experience an acceleration greater 
than that of a particle in the electric field only.

After the acceleration curve merges onto the electric force, the trajectory is the one of 
a classical particle interacting with the field only. This can, of course, only occur after the particle has emerged from the classically forbidden region.
We also repeated the same calculations using a longer wavelength ($\lambda=1568$ nm) and found 
the same general behavior, although trajectories merge more smoothly onto the electric force due to the slower variations of the barrier.
For the Coulomb case, the particle starts behaving classically as soon as the acceleration merges
onto the force $F_C(x,t)=-\partial V_C(x,t)/\partial x$, 
which includes the long-range Coulomb force. 

We also performed calculations using the OCP. The
dynamics is slightly more complicated (not shown), and
the quantum effects more pronounced, as the electron experiences
a force oriented alternatively in each direction.  The overall conclusions, however, remain the same.
The ionization occurs predominantly along the positive direction,
as several trajectories driven by the first maximum of the OCP towards $x < 0$ rescatter towards the origin.
Many of these rescattering trajectories are thrown back
at the end of the pulse towards the negative direction by
the quantum force, while other trajectories recombine in
the attractive region, perturbing bound orbits, and inducing
an ionization burst towards $x > 0$. This effect is due in
part to the fact that Bohmian trajectories cannot cross.

\begin{figure}[t]
 \begin{center}$
  \begin{array}{c}
  \vspace{-0.55cm}
\includegraphics[width=0.90\columnwidth]{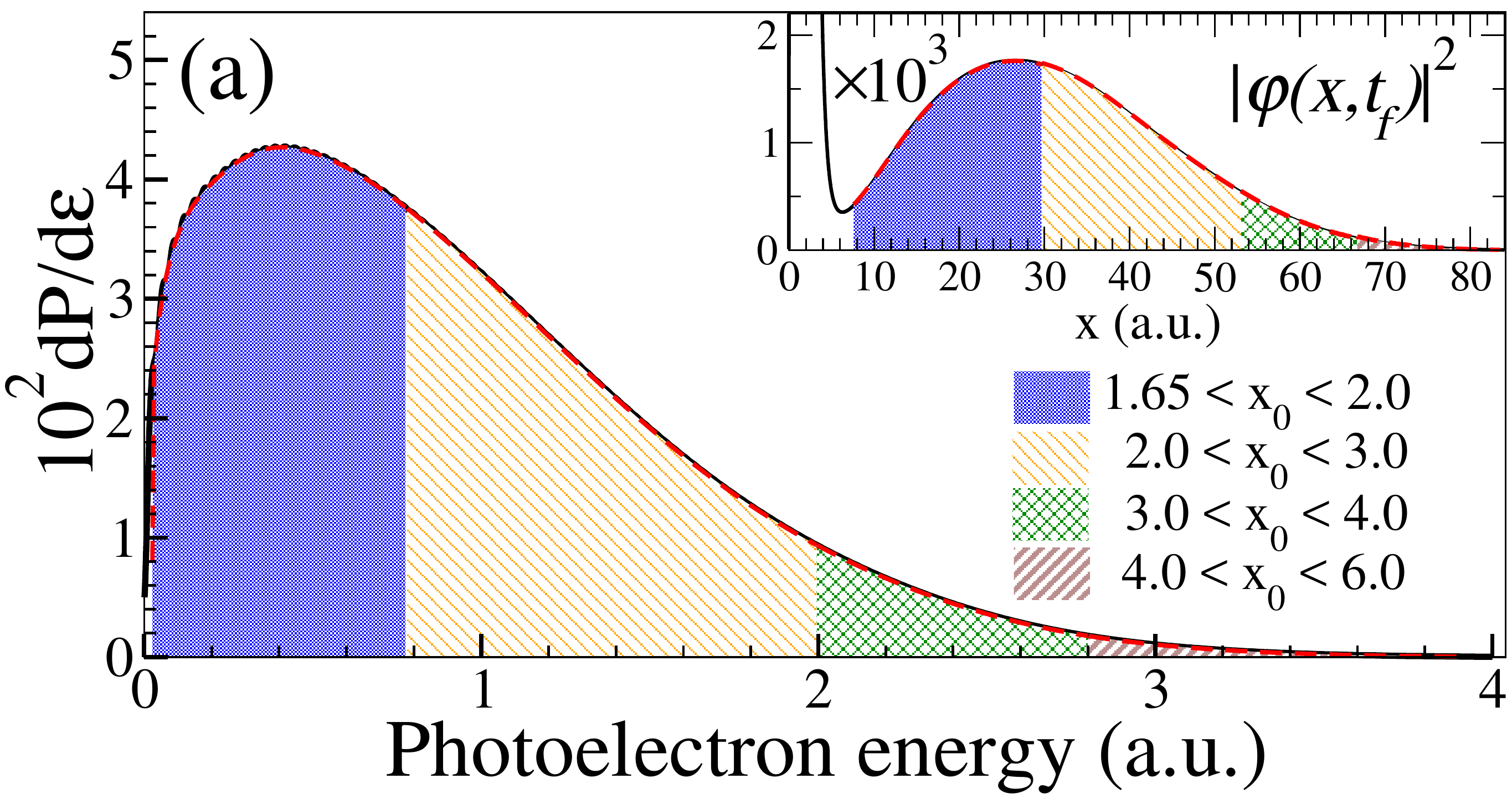} \\
\includegraphics[width=0.90\columnwidth]{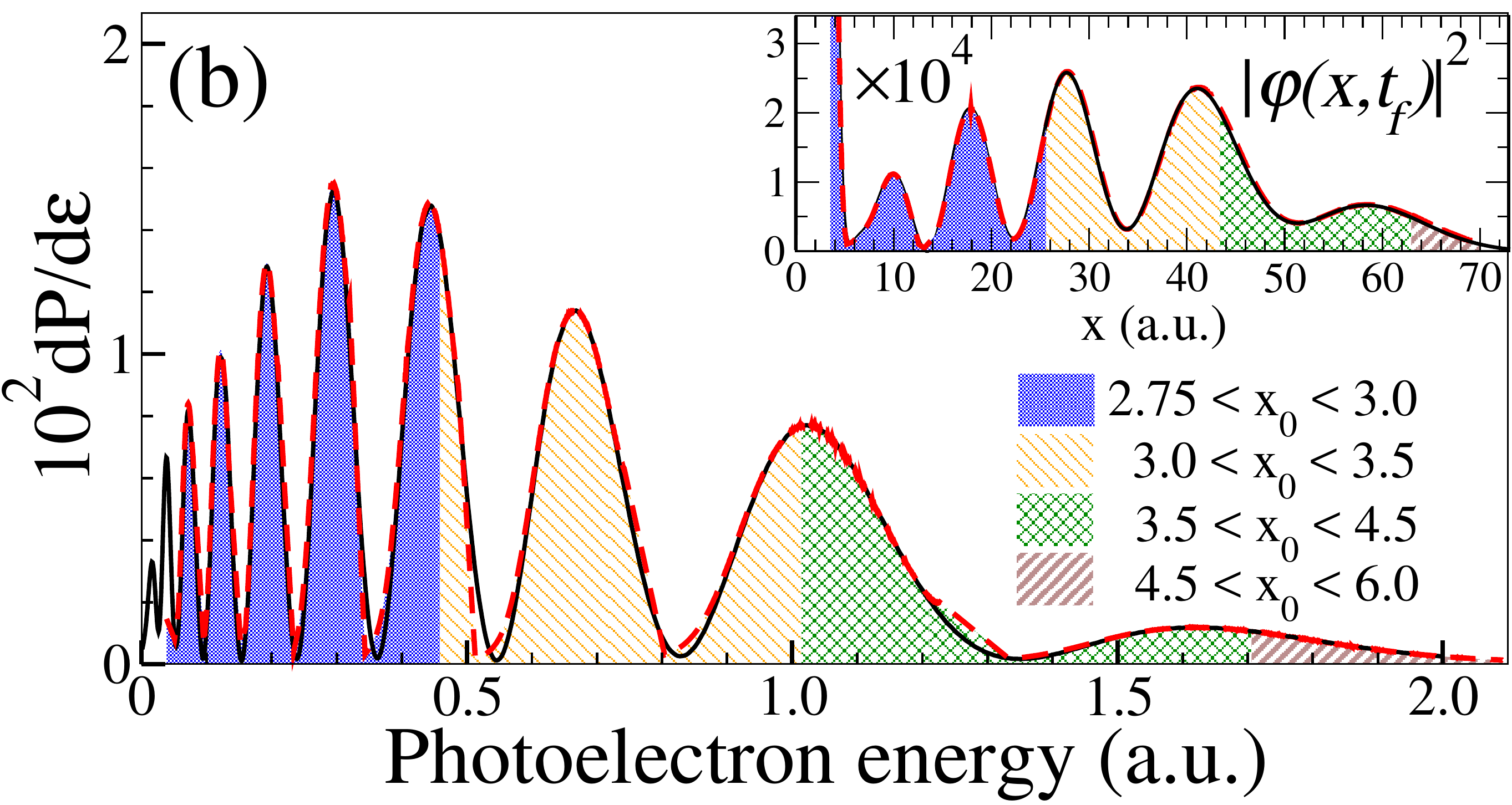}
\end{array}$
   \end{center}
 \caption{ 
 Ionization spectra for the HCP~(a) and OCP~(b) pulses calculated 
 using quantum (solid black line) and Bohmian (dashed red line) Mechanics. The insets show the
 wavefunction probability at the end of each pulse. The shaded regions give the corresponding interval
 of initial positions (see text for details).}
 \label{fig:3}
\end{figure}

The photoelectron spectrum represents a very important measurable quantity in SFI.
One additional advantage of Bohmian Mechanics is that the distribution of asymptotic speeds $v_\infty$ of ionized trajectories 
has the same density distribution $\left|\hat\varphi(p)\right|^2$ as in standard Quantum Mechanics~\cite{Durr09}. 
Here $p = v_\infty$, and $\hat\varphi$ is the 
Fourier transform of the wavefunction at $t\to\infty$. Hence, we computed the differential ionization probability 
$\Delta {\cal P}/\Delta\varepsilon=\Delta x_0\left|\varphi(x_0,0)\right|^2/\Delta\varepsilon$, 
with $\varepsilon=v_\infty^2/2$ being the asymptotic 
energy of a trajectory starting at $x_0$, and $\Delta \varepsilon$ the energy difference between nearby
trajectories starting at $x_0$ and $x_0+\Delta x_0$, respectively. 

To illustrate the ideas, we present the ionization probability associated with positive asymptotic momenta.
The results obtained in both Bohmian
and quantum approaches are plotted for both pulses in \hbox{Fig.~\ref{fig:3}} and show nearly perfect agreement.
One can thus establish a correspondence between 
the initial electron position, the region of space to where the electron has probabilistically evolved at any time, and 
the region of the ionization spectrum that its asymptotic energy will cover. 

The correspondence is illustrated in Fig.~\ref{fig:3},
where we partitioned the ionization spectrum with the associated range of initial positions,
and with the electron probability density at the end of the pulse. 
The areas labeled by the initial positions $x_1\le x_0\le x_2$ in the spectrum 
equal the probability $\int^{x(x_2;t)}_{x(x_1;t)}\left|\varphi(x,t)\right|^2dx$ to find the particle 
at any given time in this interval, and in particular the total ionization probability
${\cal P} = \int^{\infty}_{x_{\rm th}}\left|\varphi_0(x)\right|^2dx$.
Note, however, that a final momentum can in general be reached through more than one initial position for 
more complex pulses. For the OCP, the convergence of asymptotic velocities is slower, because there exist more interactions between paths.
Each peak of the ATI spectrum can be associated with a range of initial electron positions 
and the number of absorbed photons. Note that an ATI spectrum presented as histo\-grams with wide energy steps
was obtained for a one-dimensional model using Bohmian Mechanics in~\cite{Lai13}. The agreement
with the TDSE spectrum, however, was not nearly as good as the one presented here.

\subsection{Tunneling time and exit position} 
\label{sec:subsec2}
Computing the main electron ejection angle in a single-cycle circularly polarized infrared pulse,
Torlina {\it et al.}~\cite{Torlina15} claimed a zero ``tunneling ionization time''  $\uptau_{\rm ion}$ 
for an electron bound in a Yukawa
or Coulomb potential. The latter time, $\uptau_{\rm ion}=\uptau_{\rm ex}-\uptau_{\rm max}$, is defined with respect to the instant $\uptau_{\rm max}$
of maximum field, and with the exit time  $\uptau_{\rm ex}$ 
from the barrier. Ni {\it et~al.}~\cite{Ni16} 
confirmed a near-zero ionization time using classical back\-propagation in a 2D model starting with a local momentum~\cite{Feuerstein03,Wang13} 
obtained from the solution of the TDSE.  
They defined the detachment time, when $\bm{p+A}$ has no component along the field direction, as the criterion to exit the barrier. 
While their study allowed to compute the time at which the electron trajectories tunnel out of the barrier, 
it provides limited information on the dynamics preceding the tunneling. In addition, 
a recent study \cite{Camus17} showed that the electron escapes the barrier with 
nonzero longitudinal momentum, thereby contradicting the detachment criterion employed in \cite{Ni16}.
Zimmermann {\it et al.}~\cite{Zimmermann16} computed the ``tunneling'' or ``traversing time'', i.e., the time spent by the electron under the barrier, 
using standard definitions \hbox{\cite{Buttiker83,Buttiker82,Pollack84,Wigner55}}. 
Additionally, a Bohmian tunneling time to cross the barrier was computed in the adiabatic approximation
using a time-independent outgoing solution of the Schr\"{o}dinger equation in a static electric field.  
However, ignoring important aspects, such as the time variation of the barrier, and the fact that 
most of the ionization does not cross the entire barrier, lead to a drastic overestimation of the Bohmian tunneling time.
Below, we show that Bohmian Mechanics provides natural answers to the aspects mentioned above.

The Bohmian approach is well suited to define a tunneling time~\cite{Leavens90}.
In the length gauge, the tunneling condition, which includes non\-adiabatic effects, is 
given by $\varepsilon(t)=V_C(x,t)$, with $\varepsilon(t)$ denoting
the time-dependent energy of a trajectory.
In the classical case, it would correspond to a zero instantaneous speed $v(x)$ of the particle at the exit of the barrier.
Note, however, that this represents only an approximate condition, since the semi\-classical description breaks down 
at $v(x)\approx 0$. Tunneling occurs when $\varepsilon(t)<V_C(x,t)$.  This is forbidden in Classical Mechanics, but
allowed for Bohmian trajectories, as long as $\varepsilon(t)\ge V_C(x,t)+V_Q(x,t)$. The tunneling condition is 
equivalent to $T = - V_Q(x,t)$ ($T$ being the kinetic energy) and tends towards the classical limit for $V_Q(x,t)\ll1$. 
In stark contrast to the classical case, the Bohmian particle emerges from the barrier with a non\-zero velocity.

\begin{figure}[t]
\includegraphics[width=\columnwidth]{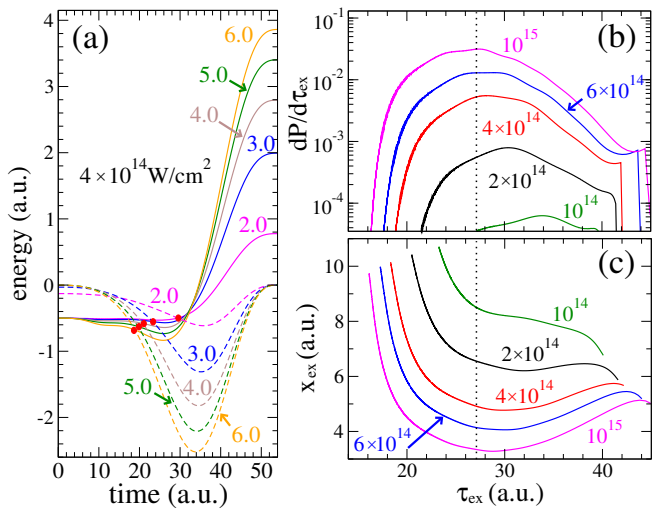}
 \caption{ 
 (a) Energy (solid lines) and $V_C(x,t)$ (dashed lines) as a function of time
 for trajectories labeled by their initial positions (the exit times $\uptau_{\rm ex}$ are marked by red dots), (b)
 probability density $d{\cal P}/d\uptau_{\rm \rm ex}$, and (c) exit point $x_{\rm ex}$ as a function of $\uptau_{\rm ex}$, 
 for several peak intensities (in W/cm$^2$). 
The vertical dotted line indicates $\uptau_{\rm max}$ (see text for details).}
 \label{fig:4}
\end{figure}

The time-dependent energy $\varepsilon(t)$ and the classical potential $V_C(x,t)$ of a few asymptotically free trajectories 
are presented in \hbox{Fig.~\ref{fig:4}~(a)} for the HCP at $4\times 10^{14}$ W/cm$^2$. 
All energy curves start at $\varepsilon_0$, have a classical asymptotic 
energy $v_\infty^2/2$ whose distribution reproduces the quantum spectrum \hbox{(see Fig.~\ref{fig:4}(a))},
and their energy averaged over all trajectories equals the total energy of the quantum system at any time \cite{Song17}.
Each energy curve crosses $V_C(x,t)$ at $\uptau_{\rm ex}$, and deviations of $\varepsilon(t)$ from $\epsilon_0$ at a crossing point are the signature 
of non\-adiabatic effects. Trajectories traveling the furthest out experience the maximum decrease in $V_C(x,t)$,
followed by the largest increase in $\varepsilon(t)$. At $I_0\ge 6\times10^{14}~$W/cm$^2$, 
trajectories with $x_{\rm th}\le x_0\le x_{\rm cl}$ cross $V_C(x,t)$ twice (not shown), 
as they enter and exit the classically forbidden region at $\uptau_{\rm en}$ and $\uptau_{\rm ex}$, respectively. 
Fitting $\uptau_{\rm ex}$ and $\uptau_{\rm en}$ as a function of the initial position,
we found, approximately, that $\uptau_{\rm ex}\propto (x_0-x_{\rm th})^{-0.2}$ 
and $\uptau_{en}\propto (x_{cl}-x_{0})^{0.3}$ for $x_{\rm th}\le x_0\le x_{\rm cl}$ 
at all intensities studied.
As expected, $\uptau_{\rm ex}\gg1$ when $x_0\approx x_{\rm th}$, while $\uptau_{\rm en}$ 
increases suddenly when $x_0$ enters the classically allowed region. Thus, 
trajectories with $x_0<x_{\rm cl}$ spend a long time inside the allowed region before entering the barrier. 

Comparing the results of Figs. \ref{fig:2}~(a) and \ref{fig:4},
we see that some trajectories become classical only a significant time after tunneling.
For example, the trajectory with initial position $x_0=6.0$ exits the barrier
at \hbox{$\uptau_{\rm ex}\approx20$ a.u.} while it becomes classical only after 30 a.u..
Moreover, in Bohmian Mechanics, the particle naturally emerges from the barrier 
with a nonzero kinetic energy $T_{\rm ex}$. In the three-dimensional case, Bohmian trajectories would
exit the barrier with a non\-vanishing longitudinal momentum.  This is in agreement with 
recent findings \cite{Camus17}, but departs from the common assumption used in both the SFA
and Ref.~\cite{Ni16}. Table~\ref{tab:1} summarizes the characteristics 
of Bohmian trajectories for different initial positions.

The differential ionization probability $d{\cal P}/d\uptau_{\rm ex}$ for trajectories 
exiting at $\uptau_{\rm ex}$ is presented in Fig.~\ref{fig:4}~(b) for several peak intensities.
The maximum exit time increases with intensity and $d{\cal P}/d\uptau_{\rm ex}$ drops sharply at large $\uptau_{\rm ex}$. 
The most probable exit time is larger than the instant of maximum field strength ($\uptau_{\rm max}\approx 27$~a.u.)
at all intensities, but tends towards $\uptau_{\rm max}$ with increasing intensity 
(the difference is $\le0.2$ a.u.\ at $1.1\times 10^{15}~$W/cm$^2$). The fact that
$d{\cal P}/d\uptau_{\rm ex}$ exhibits a broad maximum and flat regions is due to 
the sin$^2$ envelope. The maximum would appear sharper and better defined 
for a narrower envelope.   

\begin{table}[tbp]
\begin{tabular}{|p{1cm}|p{1cm}|p{1cm}|p{1cm}|}
\hline
 $x_0$ & $\uptau_{\rm ex}$&$x_{\rm ex}$&$T_{\rm ex}$\\ 
\hline\hline
$1.7$ &35.86&$5.18$&$0.050$\\
$2.0$ &29.40&$4.78$&$0.073$\\
$3.0$ &23.23&$5.55$&$0.096$\\
$4.0$ &20.79&$6.63$&$0.126$\\
$5.0$ &19.59&$7.79$&$0.159$\\
$6.0$ &18.87&$9.00$&$0.191$\\
\hline
\end{tabular}
\caption{Exit time $\uptau_{\rm ex}$, position $x_{\rm ex}$, and kinetic energy $T_{\rm ex}$ of Bohmian trajectories with different initial positions for $I_0=4\times10^{14}$W/cm$^2$.
 All values are given in a.u..}
\label{tab:1}
\end{table}

Fig.~\ref{fig:4} (c) shows the exit position $x_{\rm ex}$ from the barrier, as a function of the exit time $\uptau_{\rm ex}$, for different intensities.
The trajectories escaping the barrier the fastest have the largest exit position, since the barrier remains broad far 
from $\uptau_{\rm max}$. All curves, except for 10$^{14}$ W/cm$^2$, exhibit a minimum 
corresponding to escape when the barrier is the thinnest. Its 
position is slightly shifted from the most probable exit time and also tends towards $\uptau_{\rm max}$ with increasing intensity.
The absence of a minimum at $10^{14}~$W/cm$^2$ is due to strong non\-adiabatic effects, which allow trajectories to 
leave with $\varepsilon(t)$ significantly larger than $\varepsilon_0$, and hence with an exit position smaller 
than the adiabatic instantaneous turning point 
at the maximum field strength ($\approx 9.4$~a.u.). In fact, the energy $\varepsilon(t)$ of trajectories escaping 
with the minimum value of $x_{\rm ex}$  tends to decrease with intensity,
such that $x_{\rm ex}$ becomes larger than the adiabatic value for $I_0\ge6\times10^{14}~$W/cm$^2$.

\begin{figure}[t]
\includegraphics[width=1.0\columnwidth]{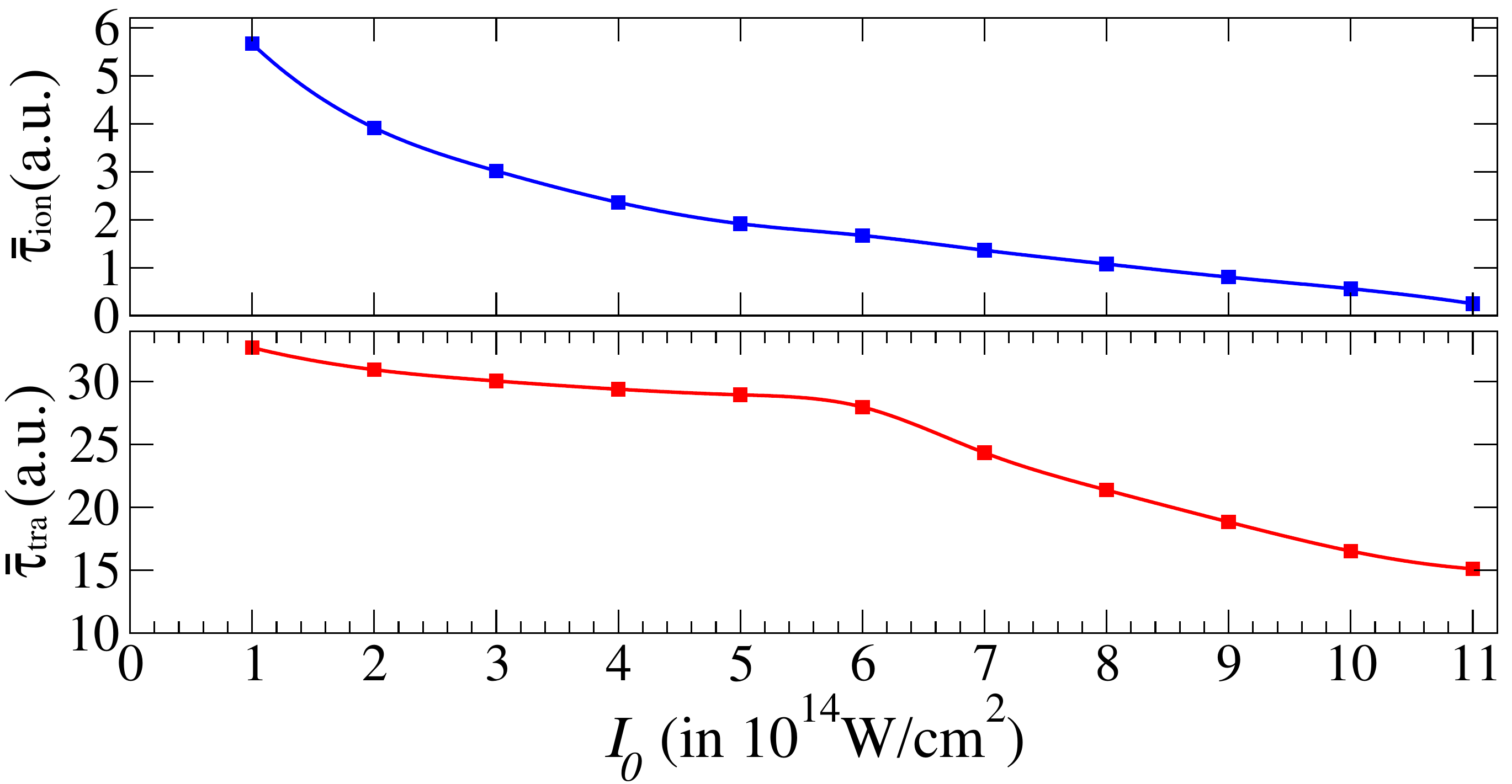}
\caption{ Mean tunneling ionization time $\bar{\uptau}_{\rm ion}$ (upper panel) and traversing time $\bar{\uptau}_{\rm tra}$ (lower panel).}
\label{fig:5}
\end{figure}

The mean values $\bar{\uptau}_{\rm ion}$ and $\bar{\uptau}_{\rm tra}$ are presented in Fig.~\ref{fig:5}, 
where  $\uptau_{\rm tra}=\uptau_{\rm ex}-\uptau_{\rm en}$
is the traversing time through the barrier. At the intensities studied, $\bar{\uptau}_{\rm ion}$ is relatively small 
and decreases towards zero as the intensity gets closer to the OBI regime. 
The variation of $\bar{\uptau}_{\rm ion}$ resembles the experimental tunneling time in~\cite{Landsman14}. 
In contrast to the results of Zimmermann {\it et al.}~\cite{Zimmermann16} who used time-independent functions,
$\bar{\uptau}_{\rm ex}$ does not become orders of magnitude larger at low intensities. 

It turns out that the traversing time shown in Fig.~\ref{fig:5} 
is a shift of 
$\bar{\uptau}_{\rm ion}$ by $\uptau_{\rm max}$ (note the different scales in the panels of Fig.~\ref{fig:5})
 until the peak intensity reaches $6\times10^{14}$ W/cm$^2$. The observed inflection point is due to contributions 
 from probability located in the inner classical region. These take a long time to reach the barrier, but they enter it 
 with a non\-negligible kinetic energy, thereby causing the traversing time to decrease faster.

\section{Conclusion}
\label{sec:conclusion}
Bohmian Mechanics possesses
many desirable features for the interpretation of complex ATI spectra 
and angular distributions with ultrashort pulses. 
Determining where the electron probability comes  from
may also provide new routes to controlling and understanding the ultra\-fast electron dynamics in molecules, for instance for
core-hole localization \cite{McCurdy17} or frustrated ionization \cite{Manschwetus09}. Using a one-dimensional model, 
we showed the potential of Bohmian Mechanics regarding the dynamics
of the electron probability over time and revealed many important features that are expected to hold in the three-dimensional case.
Once applied to a more realistic case, it might explain momentum 
distributions measured in atto\-clock experiments
by relating their different parts to tunneling time and exit position. 

\section*{Acknowledgments}
This work was supported by the 
National Science Foundation under grant \hbox{No.~PHY-1430245} and the XSEDE allocation \hbox{PHY-090031}. 
The calculations were performed on SuperMIC at the Center for Computation \& Technology at Louisiana State University.

\end{document}